\documentclass[twocolumn]{cinc}
\usepackage{graphicx,amsmath,amssymb}
\begin{document}
\bibliographystyle{cinc}
\graphicspath{{./}{figs/}}

\title{Generalization Studies of Neural Network Models for \\
Cardiac Disease Detection Using Limited Channel ECG}

\author {Deepta Rajan, David Beymer, Girish Narayan%, Charles D Coauthor$^{2}$ (note - no ``and", no periods, no degrees) \\
\ \\ % leave an empty line between authors and affiliation
 IBM Research, San Jose, CA, USA}%(note - full address at end of paper) \\
%$^2$  Second Institution, Other City, Country  }

\maketitle

% LaTeX inserts the ``Abstract'' heading in the proper style and
% sets the text of the abstract in italics as required.
\begin{abstract}

Acceleration of machine learning research in healthcare is challenged by lack of large annotated and balanced datasets. Furthermore, dealing with measurement inaccuracies and exploiting unsupervised data are considered to be central to improving existing solutions. In particular, a primary objective in predictive modeling is to generalize well to both unseen variations within the observed classes, and unseen classes.  In this work, we consider such a challenging problem in machine learning driven diagnosis – detecting a gamut of cardiovascular conditions (e.g. infarction, dysrhythmia etc.) from limited channel ECG measurements. Though deep neural networks have achieved unprecedented success in predictive modeling, they rely solely on discriminative models that can generalize poorly to unseen classes. We argue that unsupervised learning can be utilized to construct effective latent spaces that facilitate better generalization. This work extensively compares the generalization of our proposed approach against a state-of-the-art deep learning solution. Our results show significant improvements in F1-scores. 

\end{abstract}
% LaTeX inserts the extra space here automatically.

\section{Introduction}
Longitudinal patient records are central to the realm of healthcare, comprising a historical aggregation of diverse information such as diagnostic codes, lab measurements, imaging exams, text reports etc. Analyzing sequences of events and episodes are required for diagnosis and treatment planning. Harnessing the power of artificial intelligence tools and technologies for such analysis, e.g. deep learning, could potentially improve patient care quality and reduce costs by digitizing healthcare \cite{choi2016doctor}. However, the success of data-driven solutions primarily depends on two aspects: infrastructure to manage big data and deploy solutions at scale; and strong generalization abilities of computational models to produce reliable predictions in new environments. However, leveraging large volumes of patient data being routinely collected is challenged by both availability of expert annotations and our ability to discern meaningful correlations from the vast number of predictor variables. Further, these datasets are plagued by discrepancies in measurements and imbalances in disease distributions. Consequently, despite the community-wide efforts of curating representative datasets \cite{PhysioNet}, we operate in small-data regimes that result in highly biased clinical models.

In this paper, we focus on detecting cardiac abnormalities, such as myocardial infarction, a disease causing over $8$ million deaths annually \cite{ansari2017review}, using limited channel ECG. The standard $12$-channel ECG is a prevalent diagnostic modality and the primary screening exam for heart ailments, with over $300$ million signals recorded every year \cite{heden1997acute}. However, in certain cases, only a subset of these leads can be accessed. Examples include inpatient telemetry \cite{sandau2017update}, and ambulatory heart rhythm monitoring \cite{kennedy2013evolution}. In such cases, the goal is to obtain meaningful clinical conclusions using only the limited measurements~\cite{atoui2010novel}.

A popular solution for sequence modeling includes Recurrent Neural Networks (RNN) based on Long Short-Term Memory (LSTM) units, and have had proven success with clinical time-series classification \cite{choi2016doctor}. Besides, a recent empirical study \cite{bai2018empirical} deemed 1-D Convolutional Neural Networks (CNN) to be a powerful and more effective alternative to model sequences. Specifically, 1-D Residue Networks (ResNet) have become a popular solution for rhythm classification \cite{rajpurkar2017cardiologist}. More recently, new learning paradigms such as deep attention models~\cite{song2017attend} and temporal convolutional networks have been shown to produce improved performances. Broadly, these methods fall under the category of discriminatory modeling, which can be ineffective when dealing with out-of-distribution samples. Recently, Rajan \textit{et.al.} \cite{rajan2018generative} argued that a generative modeling approach can enable inference of latent features to describe complex distributions. In this paper, we build on this idea and propose a novel neural network architecture referred as \textit{ResNet++} for limited channel ECG classification. Our approach uses an unsupervised generative model to construct latent features, followed by a discriminative model ($1$-D ResNet~\cite{rajpurkar2017cardiologist}) to detect anomalies. The resulting latent feature representations implicitly exploit information from the missing channels and can predict the entire $12$-channel ECG from a subset of channels. Results show that the proposed approach provides improved generalization to new diseases, and is less sensitive to hyperparameter settings even with small and imbalanced data.

\begin{figure*}[t]
	\centering
	\centerline{\includegraphics[width=0.75\linewidth]{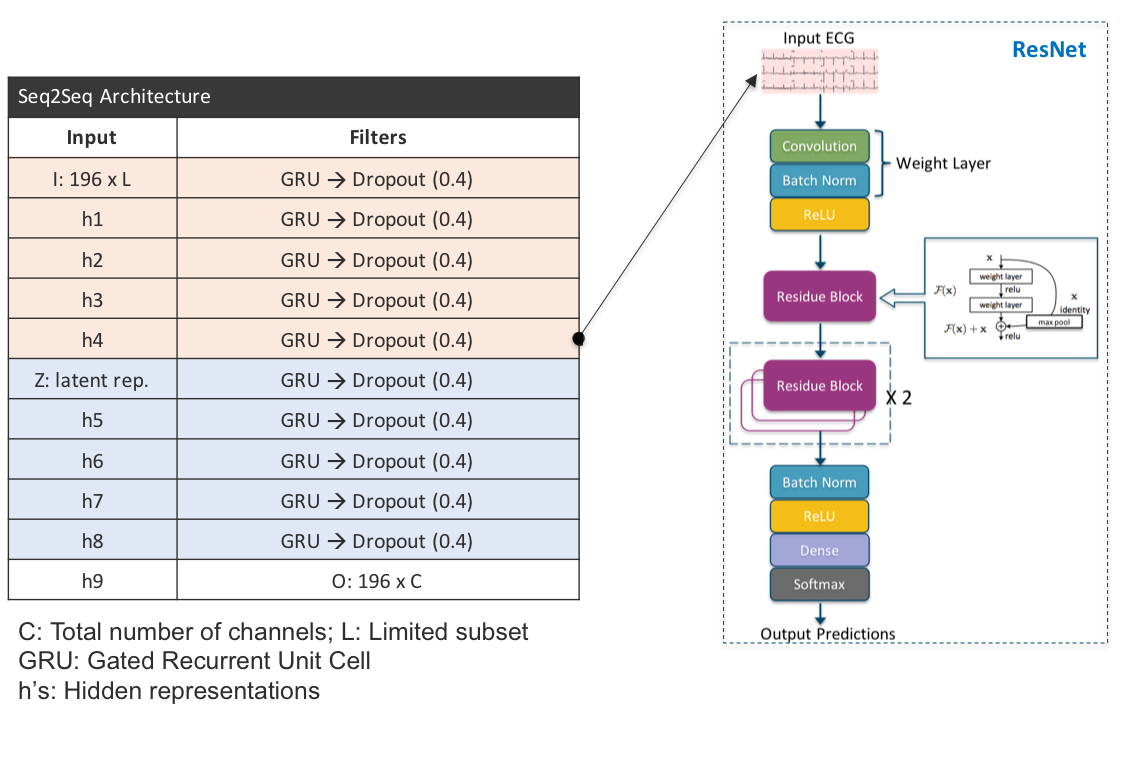}}
	\caption{Proposed ResNet++ architecture, with Stage 1 depicting the unsupervised Seq2Seq model used to construct latent representation from limited channels, and Stage 2 depicts the ResNet model for disease detection.}
	\label{fig:arch}
	\vspace{-10pt}
\end{figure*}

\section{Problem Formulation}
In this section, we provide a formal definition of the problem and introduce the notations used in the rest of this paper. In limited channel problems, ECG signals from only $1$ to $3$ leads are assumed to be available, and the goal is to build predictive models for disease detection. In this paper, we consider inferior myocardial infarcation (MI) leads (II, III, AvF) in order to build the model. From previous studies, it is known that these leads provide information to adequately localize inferior wall ischemia and infarction \cite{dubin1996rapid}. However, the ability of models trained on datasets dominated by subjects with infero MI conditions to generalize to other cardiac conditions is unknown. Such generalization studies emphasize the importance of choosing the optimal subset of leads that can be more broadly applicable in diagnosis. However, by restricting the algorithm to use only the inferior leads, disease generalization capacity of even sophisticated models suffers. To this end, we propose a novel approach that assumes access to full channel training data, builds meaningful latent spaces that can compensate for the missing channel data, and finally construct a deep neural network that performs predictions based on these latent features. Note that, in the rest of this paper, the terms leads and channels are used interchangeably. The multi-variate sequence dataset is represented as $X \in \mathbb{R}^{N \times T \times K}$, where $N$ denotes the number of training samples, $T$ denotes the number of time-steps in each measurement and $K$ indicates the total number of channels. Based on the limited channel configuration denoted by the set $\mathcal{C} =$ \{II, III, aVF\}, whose cardinality $\hat{K} \ll K$, we extract the matrix $\hat{X} \in \mathbb{R}^{N \times T \times \hat{K}}$. In order to perform implicit completion of the missing data, we propose to build a generative model that attempts to recover $X$ using $\hat{X}$. In this process, it infers a latent space that defines an effective metric to compare different samples.

\section{Proposed Approach}
\noindent \textbf{Stage 1}: As shown in Figure \ref{fig:arch}, the first stage of our algorithm employs unsupservised representation learning for predicting the complete ECG using only the 3-channel measurements. More specifically, we build an encoder-decoder architecture, commonly referred as Seq2Seq \cite{sutskever2014sequence}, with an optional attention mechanism. Though originally developed for machine translation, they are applicable to more general sequence to sequence transformation tasks. The architecture is comprised of two RNNs (based on Gated Recurrent Unit (GRU)), one each for encoder and decoder. The encoder transforms an input sequence from $\hat{X}$ into a fixed length vector, either from the last time step of the sequence or by concatenating hidden representations from all time steps. The decoder then predicts the output sequence, in our case $X$, using the encoder output.  Optionally, the decoder can also attend to a certain part of the encoder states through an attention mechanism. The attention mechanism often uses both content from the encoder states, and also context from the sequence generated so far at the decoder. Our RNNs are designed using GRU cells, which are capable of learning long-term dependencies. Each GRU cell is comprised of the following operations, implemented using fully connected networks:
\begin{align}
\nonumber &\text{(update gate):\quad} z = \sigma(x_t U^z + s_{t-1}W^z) \\
\nonumber&\text{(reset gate):\quad} r = \sigma(x_t U^r + s_{t-1} W^r)
\\
&\text{(hidden state):\quad} h = tanh(x_t U^h + (s_{t-1} \circ r )W^h)
\\
\nonumber&\text{(final output):\quad} s_t = (1-z) \circ h + z \circ s_{t-1}
\end{align}A GRU has two gates, a reset gate $r$, and an update gate $z$. While the reset gate determines how to combine the current input with the previous memory state, the update gate defines how much of the previous memory to retain as we proceed to the next step. In the simplest case, when we set the reset gate to all $1$’s and the update gate to all $0$’s it simplifies into a plain RNN model. While the underlying idea of using gating mechanisms to capture long-term dependencies is the same as in a LSTM, they are different in terms of the number of gates and the absence of an internal memory that is different from the hidden state. The generative model is trained with an $\ell_2$ loss at the decoder output. Note that, our architecture attempts to reconstruct the observed channels as well as predict the missing channel measurements.

\noindent \textbf{Stage 2}: We now design a classifier stage that exploits the latent space from the generative model trained for missing channel prediction. Interestingly, compared to discriminative models, this approach utilizes additional channel information from the training stage and builds a more effective metric for the whole data space instead of discriminating the normal/abnormal classes. Furthermore, since the first stage is unsupervised, we can use even unlabeled data to construct a more robust latent space. The classifier we use is a $1$-D ResNet architecture with convolution, ReLU, batch normalization and dropout layers, as illustrated in Figure \ref{fig:arch}.

\section{Experiment Setup}
\begin{table}[t]
	\caption{ECG class labels and sample sizes from the Physionet PTBDB used in our disease generalization study.} 
	\small
	\centering
	\renewcommand{\arraystretch}{1.3}
	\begin{tabular}{|l|l|}
		\hline
		\multicolumn{1}{|c|}{\textbf{Class Label}}                                   &\multicolumn{1}{c|}{\textbf{Sample Size}} \\
		\hline
		Myocardial Infarction: inferior               & 3222       \\
		\hline
		Myocardial Infarction: antero-septal          & 2855       \\
		\hline
		Myocardial Infarction: infero-lateral         & 1933       \\
		\hline
		Myocardial Infarction: anterior               & 1685       \\
		\hline
		Myocardial Infarction: antero-lateral         & 1603       \\
		\hline
		Myocardial Infarction: no                     & 628        \\
		\hline
		Myocardial Infarction: infero-postero-lateral & 573        \\
		\hline
		Myocardial Infarction: postero-lateral        & 185        \\
		\hline
		Myocardial Infarction: posterior              & 148        \\
		\hline
		Myocardial Infarction: infero-poster-lateral  & 111        \\
		\hline
		Myocardial Infarction: lateral                & 111        \\
		\hline
		Myocardial Infarction: infero-lateral         & 86         \\
		\hline
		Myocardial Infarction: antero-septo-lateral                          & 74         \\
		\hline
		Myocardial Infarction: infero-posterior                              & 12         \\
		\hline
		Bundle Branch Block                           & 623        \\
		\hline
		Cardiomyopathy                                & 603        \\
		\hline
		Dysrhythmia                                   & 411        \\
		\hline
		Valvular Heart Disease                        & 122        \\
		\hline
		Healthy Control                               & 3055      \\
		\hline
	\end{tabular}
\label{table:dataset}
\end{table}

In this section, we describe the dataset used for all empirical analysis that compares the performance of ResNet++ over ResNet for the problem of cardiac disease detection using limited channel ECG. We also provide details on the choice of ECG channel configurations, training parameters, and evaluation metrics employed.

\noindent \textbf{Dataset}: The widely adopted Physionet \cite{PhysioNet} repository hosts the PTB database (PTBDB) \cite{bousseljot1995nutzung} that comprises a set of $30$ seconds long, $12$ channel ECG records, sampled at $1000$ Hz. It includes ECGs collected from both healthy patients and those diagnosed with diseases such as bundle branch block, valvular heart disease etc. Every record also contains a header file with demographics, medical history and diagnosis related information curated from echocardiography exams. In our study, a total of $504$ records from $250$ patients were used. The ECG records were pre-processed to remove noise, reduce sampling rate, and divided into frames of about $3$ seconds long prior to being parsed through the neural network, as described in \cite{reasat2017detection}. The resulting dataset contains $18,040$ ECG frames downsampled to $64$ Hz . 

\noindent \textbf{ECG leads}: As depicted in Table \ref{table:dataset}, the dataset is dominated by samples representing myocardial infarction (MI) occurring in various regions of the heart, namely: inferior, anterior, lateral, septal, and posterior. These localized abnormal activities are captured in different combinations of ECG leads. For example, inferior leads (II, III, aVF) provide information to adequately localize inferior wall ischemia and infarction while anterior leads (V1, V2, V3) are effective in detecting anterior wall infarction. Similarly, lead II and V1 provide excellent assessment of atrial activity enabling detection of atrial arrhythmias, and leads V1 and V2 can often be used to differentiate left from right bundle branch block (BBB) patterns \cite{dubin1996rapid}. To study the generalization of deep models across diseases, we create two scenarios defined by choice of ECG leads which in turn determines the disease that the model can be trained to detect. A train-test pair dataset is created for each scenario, with the train data containing samples from one disease class, and the healthy class. The model is then tested on its ability to classify data it was trained to detect, in addition to the rest of the unobserved disease samples. In the first scenario, we select the leads II, III, aVF and create a training set of size $7,246$ with samples belonging to infero MI and healthy control group only, while keeping the rest of the disease classes such as posterior MI, BBB, dysrthymia etc. along with a portion of infero MI and healthy samples in the test set giving rise to a sample size $10,794$. Correspondingly, in the second scenario we select ECG leads V1, V2, V3, and create a train-test dataset pair with only antero MI and healthy classes in the training set.

\noindent \textbf{Training parameters}: Stage $1$ of ResNet++ uses $5$ GRU layers, while stage $2$ uses $7$ convolutional layers in comparison to the $13$ layers used by ResNet. Further, a batch size of $32$ samples, the Adam optimizer with a learning rate of $0.001$, and the categorical cross entropy loss function was used for training.

\noindent \textbf{Evaluation metric}: In order to evaluate the predictive models, we compute the F1-score, a summary metric typically used to trade-off between precision and recall when datasets are imbalanced.

\section{Results}
\begin{figure}[t]
	\centering
	\centerline{\includegraphics[width=0.99\linewidth]{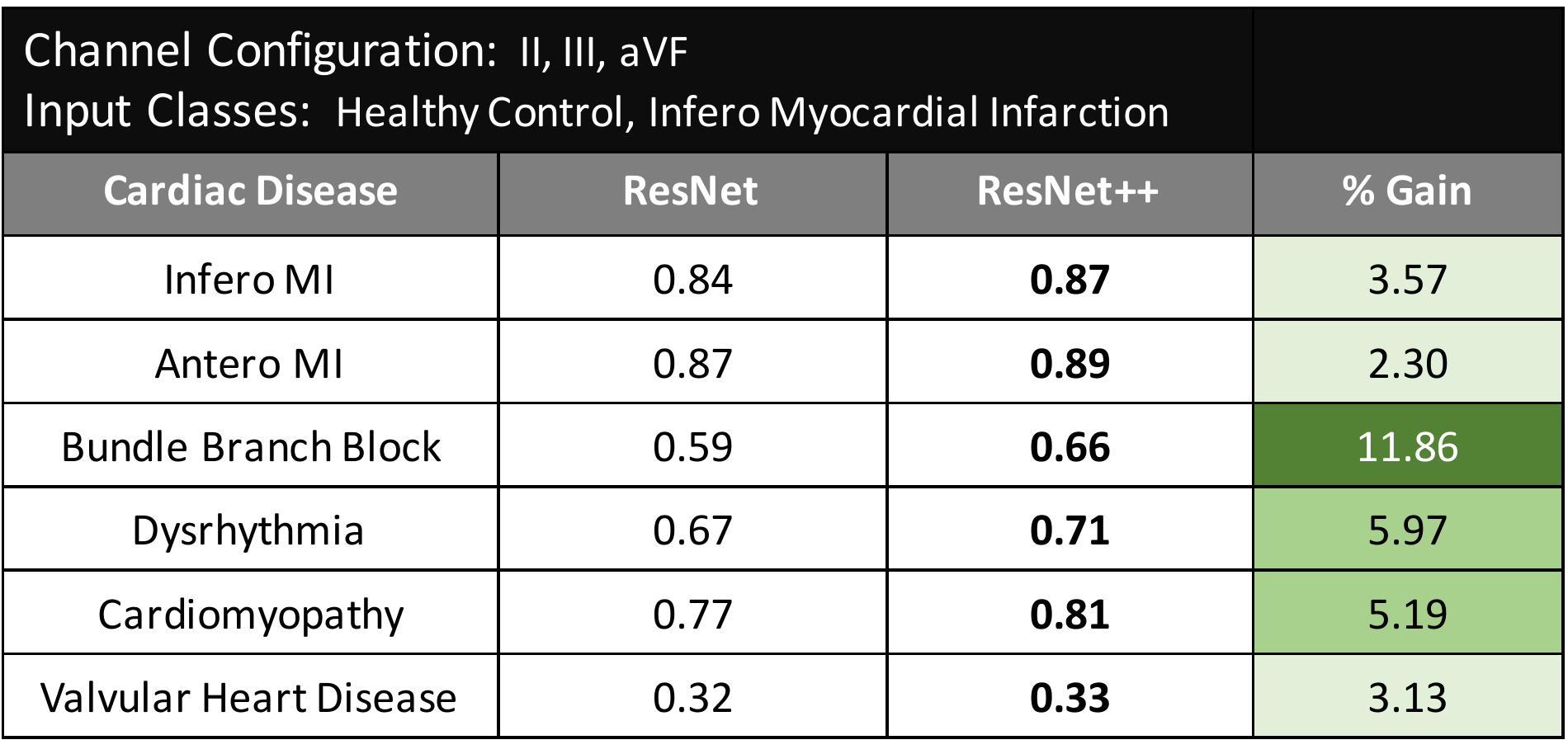}}
	\caption{F1-Scores demonstrating the improvement of ResNet++ over ResNet on the Physionet PTB dataset.}
	\label{fig:results}
	\vspace{-10pt}
\end{figure}

Our results on the PTB dataset show that, the proposed \textit{ResNet++} model trained to detect only inferior Myocardial Infarction (MI) using leads II, III, and aVF, is more effective than ResNet in detecting other MI variants (anterior, posterior, lateral), bundle branch block, dysrhythmia etc. Overall we achieve a $2$ to $11$\% improvement in F1-scores when generalizing to new unobserved diseases as shown in Figure \ref{fig:results}. More importantly, \textit{ResNet++} leverages unlabeled data in its unsupervised pre-training stage. In addition, it is robust to out-of-distribution samples, less sensitive to training hyperparameters and produces significant improvements in generalization capabilities over ResNet even in highly imbalaced data scenarios.

\bibliography{refs}

\begin{correspondence}
Deepta Rajan - 650, Harry Road, San Jose, CA, USA - 95120
\end{correspondence}

\end{document}